\documentclass[letterpaper, twocolumn, 10pt]{article}
\usepackage{usenix2020_SOUPS}

\usepackage{tikz}
\usepackage{amsmath}
\usepackage{outlines}
\usepackage[compact]{titlesec}

\begin{document}

\date{}

\title{\Large \bf Brazilian Favela Women: How Your Standard Solutions for Technology Abuse Might Actually Harm Them}


\def\plainauthor{Author name(s) for PDF metadata. Don't forget to anonymize for submission!}

\author{
{\rm Mirela\ Silva}\\
University of Florida
\and
{\rm Daniela Oliveira}\\
University of Florida
} 

\maketitle



\newcommand{\aka}{a.k.a.\ }
\newcommand{\etal}{\mbox{et al.\ }}
\newcommand{\ie}{\mbox{i.e.,\ }}
\newcommand{\eg}{\mbox{e.g.,\ }}
\newcommand{\etals}{\mbox{et al.'s\ }}

\newcommand{\Oliveira}[1]{{\todo[inline, color=purple, prepend, caption={Dr. Oliveira's comment}]{#1}}}
\newcommand{\doublecheck}{\todo[inline, color=gray, prepend, caption={Re-phrase this}]{}}
\newcommand{\needscite}{\todo[inline, color=cyan, prepend, caption={Needs citation}]{}}

\newcounter{rqcounter}
\newcommand{\researchq}[1]{
    \refstepcounter{rqcounter} 
    \noindent\textbf{(\therqcounter}) \textit{#1}
}

\begin{abstract}
Brazil is 
home to over 200M people, the majority of which have access to the Internet.
Over 11M Brazilians live in \textit{favelas}, or informal settlements with no outside government regulation, often ruled by narcos or militias. 
Victims of intimate partner violence (IPV) in these communities are made extra vulnerable not only by lack of access to resources, but by the added layer of violence caused by criminal activity and police confrontations.
In this paper, we use an unintended harms framework~\cite{Chua2020-hq} to analyze the unique online privacy needs of favela women and present research questions that we urge tech abuse researchers to consider.



\end{abstract}
\section{Introduction}

This paper focuses on the peculiarities of IPV as it intersects with technology, and the unintended harms of technology-based mitigations into a specific population: women\footnotemark[1] from Brazilian favelas. 
Brazil is the largest country in Latin America, home to over 200M people, with steep ethnoracial inequalities~\cite{noauthor_2019-nj} and wide-spread Internet access~\cite{PagBrasil2019-or, Internet_World_Stats2019-hf, noauthor_2013-vv}.
Of special note are Brazil's \textit{favelas}, \ie low-income 
urban settlements, largely characterized by high population density, informal buildings, and lack of government regulation~\cite{Santiago_Navarro_F2016-dr}, many of which are controlled by narcotraffickers or militias (for example, 850 out of the 1,025 favelas in Rio de Janeiro are not ``pacified''~\cite{Godoy2017-ug}, \ie they are unfortunately plagued by local urban violence, wherein there are constant confrontations between drug traffickers and the police).
According to the 2010 Brazilian Census~\cite{Bello2017-nw}, approximately 6\% (11.4M) of Brazilians live in favelas---while cities such as Rio de Janeiro and São Paulo (the largest city in Latin America) have 22.2\% and 11\%, respectively, of its residents living in favelas.

Brazil is also home to 40\% of all femicide in Latin America~\cite{Oas2019-eq} and ranks third in terms of global spyware usage~\cite{Nishida2019-ue, BBC_News_Brasil2019-fk}.
Due to the steep social inequalities in the country, the online privacy threats IPV victims face are not the same. 
The favela woman is made more vulnerable by the constant clash between the drug traffic that rules her community and the police forces that seek to gain power.
In this skirmish, she is left wanting for resources that could enable her peace, with very few ever seeking help from law enforcement~\cite{De_Seixas_Filho2020-tj, Santiago2019-in}, primarily because of her distrust towards the police, who either do not have the resources to counter the favela's leadership or is corrupt, enabling the narco leaders. 

Using the framework presented by Chua \etal~\cite{Chua2020-hq} to aid in the identification of unintended harms that go beyond the technology sphere, we consider the unique reality of Brazilian favelas as a case study to explore the online privacy needs of women in these communities. 
Prior to developing solutions to solve specific problems, we must first seek to understand the concerns of the problem’s target population. 
As such, this exercise is then used to present several relevant research questions that may help guide future works on online privacy protection of IPV victims in a targeted fashion, considering culture and the reality in which they are immersed.
Our goal is to join the chorus of researchers~\cite{Oliveira2017-uj,  Havron2019-xu, Redmiles2019-bq} positing that one-size-fits-all solutions might not be effective and need to consider the reality of specific vulnerable populations.


This paper is organized as follows.
Section \ref{sec:background_brazil} describes the unique characteristics of Brazilian favelas, the dire state of IPV within them, and summarizes the sparse related works on technology abuse as IPV in Brazil.
Section \ref{sec:unintendedharms} contrasts the unintended harms caused by common mitigations for tech abuse if applied to the the Brazilian favela reality, and proposes research questions.
Section \ref{sec:conclusion} concludes the paper.

\footnotetext[1]{This paper tackles IPV from the lens of female victims and male perpetrators. However, refer to RQ4 in Sec.~\ref{sec:discussion} for the limitations involving this gendered framing.}


\section{Brazilian Peculiarities and the Favela Woman}
\label{sec:background_brazil}
Favelas are informal settlements in Brazil developed due to an unmet need for affordable housing and established with no outside (\ie governmental) regulation~\cite{noauthor_2018-qe}. 
As such, society associates favelas with precarious social problems, and its residents are often highly marginalized and stigmatized.  
Approximately 2/3 of those living in favelas identify as black or brown and 60\% have suffered from some form of racial/socioeconomical prejudice 
\cite{Mello2014-bl}.
Residents of these communities also earn an average monthly income of US\$181\footnotemark[3]~\cite{Mello2014-bl}, significantly low when compared to the national average monthly income of US\$523 for white Brazilians~\cite{noauthor_2019-nj}. 
The education level of favela residents is also notable: only 27\% have completed the equivalent of a high school education~\cite{Mello2014-bl}.

A 2013 survey 
with over 750 participants in four cities (Rio de Janeiro, San Francisco, Brisbane, and London)~\cite{noauthor_undated-ky} found that, of those who had never been to a favela, 64\% had a negative perception of favela communities and 65\% held a neutral perception of the \textit{people} that live in favelas.
These numbers change dramatically with respect to people who had personally visited a favela: only 29\% had a negative perception of favelas and 71\% had a positive view of favela residents. 
Another study with 1,003 residents across 12 favelas in Rio de Janeiro~\cite{Mello2014-bl} found that 94\% of residents considered themselves happy and optimistic, and a whopping 2/3 would not move out of a favela, even if their income doubled.

Even so, favelas are very much a product of their environment. 
This section illustrates how the high rates of IPV coupled with the proliferation of internet usage makes Brazil, and especially its favelas, worthy of focused research into technology abuse. 


\footnotetext[3]{Conversion rate used: US\$1 to R\$5.35.}
\subsection{An Epidemic of Violence}
\label{sec:ipvBRBRBR}
Domestic violence in Brazil is an epidemic.
According to a 2015 report 
\cite{Waiselfisz2015-zs}, 
Brazil ranked fifth in terms of highest worldwide femicide rate.
Most alarmingly, a 2014 study 
\cite{Avon2014-ck} estimates that 66\% of young Brazilian women (ages 16--24) have suffered from some form of violence or control from an intimate partner.
Forty-three percent of this sample also confessed to having witnessed their mother being abused (either verbally or physically) by her partner.
 
Between 2003 and 2013, the femicide rate 
rose by 19.5\% for black and brown women~\cite{Waiselfisz2015-zs} (the predominant races of favela dwellers).
In 2013, there were only 500 Delegacias da Mulher (specialized police departments responsible for handling IPV) in Brazil~\cite{noauthor_2013-xm} and in 2019, only 2.4\% of Brazilian cities were home to women's shelters~\cite{2019-xl}.
\textbf{The favela victim of IPV is even more vulnerable and fragile compared to the average Brazilian as she is extremely limited in her access to resources that would allow her to escape an IPV situation}~\cite{noauthor_2014-qn, Santiago2019-in, Peixoto2017-kv}.

de Seixas Filho \etal~\cite{De_Seixas_Filho2020-tj} interviewed 80 women in the Complexo do Alemão, one of the largest favelas in Rio de Janeiro, and concluded that this community had both alarmingly high rates of violence against women and scarce services for them. 
Of the women interviewed, 85\% reported having suffered IPV, but only 10\% went on to fully report their cases to the police, due to lack of information,
fear of her aggressor, and/or lack of faith in the justice system.
In 91\% of these IPV cases, their aggressor was a current or former partner.

Brazilian favelas, however, are not uniformly the same.
The Complexo da Maré, home to 16 favela communities and over 140,000 residents, is not pacified (\ie they face constant violence between the invading police forces and the criminals in power). 
A recent study interviewed staff workers at a Reference Center for Women's Care (CRAM) located in the Maré~\cite{Santiago2019-in} and found that the goal of the women who suffer IPV in this community was solely the end of their partners' abuse towards them, without 
ending their relationship with their partners (due to financial dependence or simply love) nor involving law enforcement.
Not only did some of the victims have little faith in the justice system, but they also feared that the police would invade the Maré on their behalf---risking retaliation from narcos, especially since many of their partners were themselves involved in the drug trade.
As such, escaping to shelters or using the justice system are only seen as last-case scenarios; instead, the women often resort to asking the drug gangs for help 
in the hopes that ``correctional measures'' outside of the Brazilian criminal justice system will be taken towards their abusers.
The authors emphasize that these highly complex scenarios 
have no easy or quick solutions and attempting to simplify the problem could instead exacerbate it.
As such, \textbf{the vulnerability of women in favelas is further entangled in their constant violations of rights by their partners, the State, or the drug traffickers that rule their communities}.
Abuse that is not physical in nature (\eg threats, harassment, or humiliation) are rarely seen in Brazil as a form of abuse by both men~\cite{noauthor_2013-xm} and women~\cite{De_Seixas_Filho2020-tj, Santiago2019-in} alike. 
This is especially alarming as moral and psychological violence are oftentimes precursors to physical violence in IPV situations~\cite{noauthor_2013-xm, Mohandie2006-gv, CDC_NISVS_2010}.
de Seixas Filho \etal conclude that providing women with knowledge and information is an urgent need in favelas, calling for the use of technology and the internet as the only ways to reach many favela women.

\subsection{A Plugged-In Society}
\label{sec:techuse}
Despite these social inequalities, the use of internet-connected devices is nearly universal in Brazil.
Among adults ages 18--55, 92\% own or have access to smartphones;
this rate varies among generations, but 
ages 18--34 take the lead, wherein a whopping 85\% own a smartphone~\cite{PagBrasil2019-or}.
(This is especially concerning given that tech abuse victimization is highest among women under the age of 25, as detailed in Sec.~\ref{sec:background_stalking}.)

SMS texting was never accessible for the average Brazilian, as the cost of SMS was as much as 55 times higher than in North America, a price far too expensive for most Brazilians~\cite{Saboia2016-bc}.
This problem was solved with the advent and popularization of WhatsApp, allowing people to send messages for free.
By 2016, 96\% of Brazilians who owned a smartphone used WhatsApp as their primary method of communication~\cite{Saboia2016-bc}.
Now, WhatsApp is not only a personal communication medium for Brazilians, but used by (small and large) businesses and government agencies for communication.
For example, it is common for Brazilians to schedule appointment to services or for medical doctors to communicate with their patients informally via WhatsApp.
These rates all reflect Brazil's ranking of the world's fifth highest number of Internet users~\cite{Internet_World_Stats2019-hf}.

Although favelas are markedly of lower SES, they nonetheless are still highly connected communities. 
A 2013 study conducted in 63 favelas in several metropolitan cities in Brazil~\cite{noauthor_2013-vv} found that 9 out of 10 favela residents owned a smartphone, and half had access to the Internet---a number that rises to 78\% for those between the ages of 16 and 29.
Of note, 1 in 4 residents admit to sharing their WiFi with a neighbor due the favelas' culture of sharing.

\subsection{An Unstudied Enigma}
\label{sec:background_stalking}


In October of 2019, Brazil ranked third in terms of global spyware usage~\cite{Nishida2019-ue, BBC_News_Brasil2019-fk}. 
But research that touches on technology abuse perpetrated by intimate partners in Brazil is scant. 
It is slightly easier to find research in Brazil on related topics, \eg \textit{stalking victimization}, which can involve real-world contact (\eg physically following a victim or showing up to their workplace) and/or contact in the online domain (\eg sending unwanted texts or tracking someone's location). 

In this vein, nonprofit Instituto Avon 
\cite{Avon2014-ck}, found that 32\% of young Brazilian women (ages 16--24) have had an intimate partner check their email or social media without their permission, and 15\% were forced to reveal their Facebook passwords to their partner.
Boen and Lopes~\cite{Boen2019-xc} interviewed 205 university students in Campinas (located in the state of São Paulo) and found that most victims of stalking were women, with 87.5\% of stalkers being someone known to the victim. 
Alarmingly, 50.4\% of the students surveyed reported experiencing a stalking incident---a significantly higher number compared to the 16.7\% reported by the CDC 
\cite{CDC_NISVS_2010}. 
This could be due to the sample's average age of 24.7, which can nonetheless highlight that stalking is more prevalent amongst the young (also supported by the CDC). 
But as the authors acknowledge, this study was conducted on a small and specific subgroup of students from a private university.
It can be reasonably assumed that the students were of relatively high SES compared to the Brazilian favela population, due to access to both higher and privatized education.

Security and privacy research in Brazilian favelas is nearly nonexistent, with Arora and Scheiber~\cite{Arora2017-hf} as a clear exception. 
The authors used a myriad of ethnographic methods on 22 Brazilian and 22 Indian youths living in favela-like dwellings or peri-urban areas. 
They found that all youths struggled to define privacy, while the Brazilian participants were more cautious and distrustful of the Internet than their Indian counterparts.
Yet in the context of relationships, the Brazilians nonetheless were comfortable with sharing their cellphone PIN/password with their partners.
This mentality is supported by~\cite{Avon2014-ck}, wherein 51\% of participants
revealed that they share the PIN/password for their mobile devices with their partners. 
Although a good first step, the context of technology abuse as a form of IPV was outside the scope of Arora and Scheiber and therefore remains an unstudied enigma.

\section{The Privacy Needs of the Favela Woman}
\label{sec:unintendedharms}

Chua \etal~\cite{Chua2020-hq} proposed a framework for identifying a series of unintended real-world harms of countermeasures in cybersecurity,
summarized into seven broad categories: 
(1) \textbf{Displacement} of harm onto others;
(2) \textbf{Insecure Norms} that encourage potentially greater harm;
(3) \textbf{Additional Costs} that burden the stakeholder;
(4) \textbf{Misuse} of the countermeasure to create new harms;
(5) \textbf{Misclassification} that can incorrectly label non-malicious content/individuals as malicious;
(6) \textbf{Amplification} causing an increase of the very behavior targeted for prevention; and
(7) \textbf{Disruption} of other or more effective countermeasures. 
We use this framework to exemplify and emphasize the idea that the favela woman's reality comes with a plethora of nuanced and delicate considerations that the cybersecurity community must consider whilst developing solutions to related problems.
Tech abuse as a form of IPV is still an open problem with no best practices available~\cite{Jee2019-tv}; as such, the countermeasures detailed below are not exhaustive.

\noindent
\textbf{Check the suspect device's privacy settings.}
Although data is not specific to favelas, Brazil is listed among countries where less than half of the population possesses basic computer skills (\eg copy and paste, sending email with attachment)~\cite{noauthor_2019-ou}. 
This countermeasure may therefore \textit{displace harm} to individuals without computer or privacy literacy, such as young Brazilians in favelas~\cite{Arora2017-hf}.
This could also create an \textit{insecure norm} by giving victims a false sense of privacy. 
WhatsApp Web, for example, allows WhatsApp to be used on other devices, detectable only if the victim were to routinely monitor the app's settings for the addition of suspicious devices.

\noindent
\textbf{Attempt to leave or move away from abusive partner.}
This yet again \textit{displaces harm} to the favela woman, who may have nowhere to escape; not only do most people in favelas prefer to remain in their communities~\cite{Mello2014-bl}, but the favela woman in IPV situations will only consider leaving her partner or fleeing to a shelter in worst-case scenarios~\cite{De_Seixas_Filho2020-tj, Santiago2019-in}. 
Moving away would also come with \textit{additional costs} 
of possibly tremendous psychological, financial, and social repercussions. 
This act may also trigger a \textit{misuse} of the countermeasure, placing the favela woman in greater harm if her partner is a member of the favela's narco network, wherein escaping could be used as a reason for further abuse (\eg losing contact with her children, gang revenge, social isolation from other favela residents).

\noindent
\textbf{Seek police help.}
This countermeasure has an inherent \textit{misclassification} between what the police and the favela woman believe will be the best solution; as such, IPV victims in favelas rarely report their abuser to the police 
\cite{De_Seixas_Filho2020-tj, Santiago2019-in}. 
This may also cause an \textit{amplification} of violence, as police are simply not welcome in favelas, and 
may incur the ire of narcos in the favela, risking not only her life, but also coming at the \textit{additional cost} of social isolation from other residents.

\noindent
\textbf{Decrease technology use.}
As explained in Sec.~\ref{sec:techuse}, WhatsApp is ingrained into the Brazilian lifestyle. 
Isolation from WhatsApp means losing her support circle and would be \textit{misused} by the abuser to further isolate his victim.
This could also come with the \textit{additional cost} of loss of income as 47\% of favela residents have ``informal jobs'' 
\cite{Mello2020-gp}, and access to WhatsApp to communicate with clients 
is pivotal. 
Internet has also become necessary in favelas to study and seek an education, and of the 
low SES Brazilians that have access to the Internet, 85\% access the Internet exclusively through their smartphones~\cite{Saboia2020-kw}.
Therefore, this could \textit{disrupt} other countermeasures, as removing access to the Internet would further isolate victims from opportunities of growth and help.

\noindent
\textbf{Replace or discard suspect device(s).}
Similarly, 
low SES favela women that are financially dependent on their abuser would not be able to incur this \textit{additional cost} nor might it be worth the risk of \textit{amplification} of violence as retaliation. 

\noindent
\textbf{Do not share phone with others.}
This would \textit{displace} harm onto the favela woman, as there is a culture of sharing within favela communities~\cite{Mello2014-bl, Saboia2020-kw, Arora2017-hf, Avon2014-ck}.
Smartphones are also sometimes shared within families as the only way to access the Internet~\cite{Saboia2020-kw}, and \textit{misclassifying} who she cannot trust with her phone could lead to the \textit{additional cost} of losing some in her social support circle or an \textit{amplification} of violence. 

\subsection{Discussion and Research Questions}
\label{sec:discussion}
Although other works~\cite{Chatterjee2018-kr, Matthews2017-qy, Freed2018-ax} have been equally critical of these aforementioned mitigations, research also shows that privacy and security solutions must be targeted~\cite{Oliveira2017-uj, Chua2020-hq, Woodlock2017-hq, Havron2019-xu, Redmiles2019-bq}. 
As such, this exercise raised several questions that can help guide future work on technology abuse in IPV scenarios:

\researchq{What available tools or mitigations can be leveraged for the favela woman?}
Examples such as EasyTro~\cite{Horne2020-jl}, which helps U.S. IPV survivors file temporary retraining orders, or Havron \etal's~\cite{Havron2019-xu} clinical computer security approach to provide personalized technology abuse help from technologists, can be starting points, but must be heavily adapted to fit the needs of the favela woman.
Technology can also create ``invisibility cloak'' capabilities for communication and location privacy that does not raise red flags for abusers and, if confronted, gives the victim plausible deniability.
Ideas include adding random noise to real location in a way that is believable (\eg \cite{Narain2019-uq}), or the implementation of ephemeral (Snapchat-like) communication in WhatsApp without having to resort to known ephemeral apps.

\researchq{How can information be disseminated to affect change within the very populations we aim to help?} Women in favelas want to be informed, and are able and willing to do so via technology~\cite{De_Seixas_Filho2020-tj}, but many have low education levels~\cite{Mello2014-bl}, low privacy literacy~\cite{Arora2017-hf}, and lack an awareness of their own rights~\cite{De_Seixas_Filho2020-tj, Santiago2019-in}. 
Practical guides that break technical solutions down to fine grained levels while using language appropriate for the average favela woman's educational level are needed.

\researchq{What biases are researchers bringing to the tech abuse discussion?}
As noted in Freed \etal~\cite{Freed2019-gu}, we, as researchers, must navigate our own biases as the IPV survivor is the ultimate authority on their situation.
Outsiders have an inherently negative bias of favelas~\cite{noauthor_undated-ky}, and it may be more difficult for us to be welcomed into favelas and earn the trust of victims in self-reporting their perceptions.
Without researchers' access to their reality, effective solutions cannot be developed or evaluated.
We must work within the limitation that the favela woman is unlikely to leave her abuser or the favela 
\cite{De_Seixas_Filho2020-tj, Santiago2019-in}.
Several works~\cite{Arora2017-hf,noauthor_2013-xm, De_Seixas_Filho2020-tj, Santiago2019-in} also address that the domestic sphere in Brazilian society is tethered to the patriarchy with many women complicit in it. 
Monetary costs are undoubtedly a factor, but many of the additional costs described in Sec.~\ref{sec:unintendedharms} focus on the impacts of social isolation, and as such, we cannot undervalue the social component of the favela.

\researchq{What other vulnerable populations are in the favela?} 
This paper focused on 
tech abuse 
from a lens of women as its primary victim. 
But it is also imperative to recognize that men as victims are largely under-reported and underrepresented in this context.
This mindset can also be heteronormative, leading to the exclusion of the LGBTQ+ community, which, in Brazil, faces high victimization rates of discrimination, violence, and homicide~\cite{Jacobs2016-fy}. 

\researchq{How exactly is technology abused in favelas?} 
Much of the works cited about IPV in favelas were conducted in Rio de Janeiro, and problems faced in Rio may not necessarily transfer into other regions of Brazil. 
Sec.~\ref{sec:background_stalking} also focused largely on the problems of stalking victimization and online harassment.
Cyberbullying and revenge porn, for example, might be promising research interests for favelas.
Arora and Scheiber \cite{Arora2017-hf}, for example, found that Brazilian youths from favelas held views that women victim of revenge porn are at fault for their own victimization. 

\researchq{How can research in favelas be propagated into other communities?}
Favelas are not necessarily ``unique'' to Brazil; informal settlements can be found in several countries all over the world. 
While they all share commonalities that can undoubtedly be leveraged, 
each country or region's socio-political background is different. 
Solutions and mitigations that help ameliorate the problem in these regions will undoubtedly provide insight and innovations focused on privacy, with great potential of benefiting other ethnic groups in the U.S. and abroad.


\section{Conclusion}
\label{sec:conclusion}
This paper aimed to show that technology abuse in Brazilian favelas, though highly under-reported and not fully studied, is an issue worthy of attention.
The presence of high rates of femicide and intimate partner violence (IPV), coupled with high rates of Internet use is on par with the requirements of technology abuse.
But public and academic awareness of this issue is still scarce.
Via the use of an unintended harms framework, we highlighted the unique vulnerabilities of the favela woman in scenarios of IPV and technology abuse, and suggested research questions to help guide future works.

\bibliographystyle{plain}
\bibliography{bib/references.bib}

\begin{thebibliography}{10}

\bibitem{noauthor_undated-ky}
{Perceptions Survey}.
\newblock Technical report, {Catalytic Communities (CatComm)}.
\newblock Accessed: 2020-5-16.

\bibitem{noauthor_2013-vv}
{Metade dos moradores das favelas tem acesso {\`a} internet}.
\newblock
  \url{https://epocanegocios.globo.com/Informacao/Resultados/noticia/2013/10/metade-dos-moradores-das-favelas-tem-acesso-internet.html},
  October 2013.
\newblock Accessed: 2020-5-16.

\bibitem{noauthor_2013-xm}
{Percep{\c c}{\~o}es dos homens sobre a viol{\^e}ncia dom{\'e}stica contra a
  mulher}.
\newblock Technical report, Instituto Avon, Data Popular, November 2013.

\bibitem{noauthor_2014-qn}
{Mulheres sofrem viol{\^e}ncia no asfalto ou na favela - Geled{\'e}s}.
\newblock \url{https://www.geledes.org.br/asfalto-favela-violencia-de-genero/},
  August 2014.
\newblock Accessed: 2020-5-16.

\bibitem{Avon2014-ck}
{Viol{\^e}ncia contra a mulher: o jovem est{\'a} ligado?}
\newblock Technical report, Instituto Avon, 2014.

\bibitem{noauthor_2018-qe}
{Sustainable Favela Network: Map (2017)}.
\newblock Technical report, \{Catalytic Communities (CatComm)\}, February 2018.

\bibitem{noauthor_2019-nj}
{Desigualdades Sociais por Cor ou Ra{\c c}a no Brasil}.
\newblock Technical report, IBGE, 2019.

\bibitem{noauthor_2019-ou}
{Measuring digital development: Facts and figures}.
\newblock Technical report, International Telecommunication Union, 2019.

\bibitem{Arora2017-hf}
Payal Arora and Laura Scheiber.
\newblock Slumdog romance: Facebook love and digital privacy at the margins.
\newblock {\em Media, Culture \& Society}, 39(3):408--422, April 2017.

\bibitem{BBC_News_Brasil2019-fk}
{BBC News Brasil}.
\newblock {'Stalkerware': o software que espiona seu parceiro cada vez mais
  usado no Brasil}.
\newblock {\em BBC}, October 2019.

\bibitem{Bello2017-nw}
Luiz Bello.
\newblock {Dia Nacional da Habita{\c c}{\~a}o: Brasil tem 11,4 milh{\~o}es de
  pessoas vivendo em favelas}.
\newblock
  \url{https://agenciadenoticias.ibge.gov.br/agencia-noticias/2012-agencia-de-noticias/noticias/15700-dados-do-censo-2010-mostram-11-4-milhoes-de-pessoas-vivendo-em-favelas},
  August 2017.
\newblock Accessed: 2020-5-15.

\bibitem{CDC_NISVS_2010}
{Black, M.C., Basile, K.C., Breiding, M.J., Smith, S.G., Walters, M.L.,
  Merrick, M.T., Chen, J., \& Stevens, M.R.}
\newblock The national intimate partner and sexual violence survey ({NISVS)}:
  2010 summary report.
\newblock Technical report, National Center for Injury Prevention and Control,
  Centers for Disease Control and Prevention, Atlanta, GA, 2011.

\bibitem{Boen2019-xc}
Mariana~Tordin Boen and Fernanda~Luzia Lopes.
\newblock {Vitimiza{\c c}{\~a}o por stalking: um estudo sobre a preval{\^e}ncia
  em estudantes universit{\'a}rios}.
\newblock {\em Revista Estudos Feministas}, 27(2), 2019.

\bibitem{Chatterjee2018-kr}
R~Chatterjee, P~Doerfler, H~Orgad, S~Havron, J~Palmer, D~Freed, K~Levy, N~Dell,
  D~McCoy, and T~Ristenpart.
\newblock {The Spyware Used in Intimate Partner Violence}.
\newblock In {\em {2018 IEEE Symposium on Security and Privacy (S\&P)}}, pages
  441--458, May 2018.

\bibitem{Chua2020-hq}
Yi~Ting Chua, Simon Parkin, Matthew Edwards, Daniela Oliveira, Stefan
  Schiffner, Gareth Tyson, and Alice Hutchings.
\newblock {Identifying Unintended Harms of Cybersecurity Countermeasures}.
\newblock In {\em {2019 APWG Symposium on Electronic Crime Research (eCrime)}},
  2020.

\bibitem{2019-xl}
{Daiane Costa, Constança Tatsch}.
\newblock {Treze anos ap{\'o}s Lei Maria da Penha, s{\'o} 2,4\% das cidades
  t{\^e}m casas-abrigo para mulheres}.
\newblock
  \url{https://oglobo.globo.com/sociedade/celina/treze-anos-apos-lei-maria-da-penha-so-24-das-cidades-tem-casas-abrigo-para-mulheres-23972179},
  September 2019.
\newblock Accessed: 2019-11-30.

\bibitem{De_Seixas_Filho2020-tj}
Jos{\'e}~Teixeira de~Seixas~Filho, Flavia~Miranda de~Freitas~Xerfan,
  {S{\'\i}lvia Concei{\c c}{\~a}o Reis}, and Patr{\'\i}cia~Maria Dusek.
\newblock An{\'a}lise da violenl{\^e}ncia dom{\'e}stica no amviente da favela.
\newblock {\em Revista Valore}, 5(0):5013, April 2020.

\bibitem{Freed2019-gu}
Diana Freed, Sam Havron, Emily Tseng, Andrea Gallardo, Rahul Chatterjee, Thomas
  Ristenpart, and Nicola Dell.
\newblock ``{I}s my phone hacked?'' analyzing clinical computer security
  interventions with survivors of intimate partner violence.
\newblock In {\em Proceedings of the ACM on Human-Computer Interaction (PACM
  HCI)}, volume~3, pages 202:1--202:24, New York, NY, USA, November 2019. ACM.

\bibitem{Freed2018-ax}
Diana Freed, Jackeline Palmer, Diana Minchala, Karen Levy, Thomas Ristenpart,
  and Nicola Dell.
\newblock {``A Stalker's Paradise'': How Intimate Partner Abusers Exploit
  Technology}.
\newblock In {\em {Proceedings of the 2018 CHI Conference on Human Factors in
  Computing Systems}}, number Paper 667 in CHI '18, pages 1--13, New York, NY,
  USA, April 2018. Association for Computing Machinery.

\bibitem{Godoy2017-ug}
Roberto Godoy and Roberta Pennafort.
\newblock {Rio tem 850 comunidades controladas por traficantes ou milicianos,
  diz estudo}.
\newblock
  \url{https://noticias.uol.com.br/ultimas-noticias/agencia-estado/2017/09/27/policias-e-forcas-armadas-mapeiam-850-areas-no-rio-sob-dominio-do-crime.htm},
  September 2017.
\newblock Accessed: 2020-5-26.

\bibitem{Havron2019-xu}
Sam Havron, Diana Freed, Rahul Chatterjee, Damon McCoy, Nicola Dell, and Thomas
  Ristenpart.
\newblock {Clinical computer security for victims of intimate partner
  violence}.
\newblock In {\em {28th {USENIX} Security Symposium ({USENIX} Security 19)}},
  pages 105--122. USENIX Association, 2019.

\bibitem{Horne2020-jl}
Lydia Horne.
\newblock {Coders Who Survived Human Trafficking Rewrite Their Identities}.
\newblock {\em Wired}, April 2020.

\bibitem{Internet_World_Stats2019-hf}
{Internet World Stats}.
\newblock {Internet Top 20 Countries With the Highest Number of Internet
  Users}.
\newblock \url{https://www.internetworldstats.com/top20.htm}, June 2019.
\newblock Accessed: 2019-12-7.

\bibitem{Jacobs2016-fy}
Andrew Jacobs.
\newblock Brazil is confronting an epidemic of anti-gay violence.
\newblock {\em The New York Times}, July 2016.

\bibitem{Jee2019-tv}
Charlotte Jee.
\newblock {How ``stalkerware'' apps are letting abusive partners spy on their
  victims}.
\newblock {\em MIT Technology Review}, July 2019.

\bibitem{Matthews2017-qy}
T.~{Matthews}, K.~{O’Leary}, A.~{Turner}, M.~{Sleeper}, J.~P. {Woelfer},
  M.~{Shelton}, C.~{Manthorne}, E.~F. {Churchill}, and S.~{Consolvo}.
\newblock Security and privacy experiences and practices of survivors of
  intimate partner abuse.
\newblock {\em IEEE Security and Privacy}, 15(5):76--81, 2017.

\bibitem{Mello2020-gp}
Daniel Mello.
\newblock {Covid-19: 70\% dos moradores de favelas tiveram redu{\c c}{\~a}o da
  renda}.
\newblock
  \url{https://agenciabrasil.ebc.com.br/economia/noticia/2020-03/covid-19-70-dos-moradores-de-favelas-tiveram-reducao-da-renda},
  March 2020.
\newblock Accessed: 2020-5-26.

\bibitem{Mello2014-bl}
K{\'a}thia Mello.
\newblock {Com 2 milh{\~o}es de moradores, favelas do Rio seriam 7ª maior
  cidade do pa{\'\i}s}.
\newblock
  \url{http://g1.globo.com/rio-de-janeiro/noticia/2014/09/com-2-milhoes-de-moradores-favelas-do-rio-seriam-7-maior-cidade-do-pais.html},
  September 2014.
\newblock Accessed: 2020-5-16.

\bibitem{Mohandie2006-gv}
Kris Mohandie, J~Reid Meloy, Mila~Green McGowan, and Jenn Williams.
\newblock {The RECON typology of stalking: reliability and validity based upon
  a large sample of North American stalkers}.
\newblock {\em Journal of Forensic Sciences}, 51(1):147--155, January 2006.

\bibitem{Narain2019-uq}
Sashank Narain, Aanjhan Ranganathan, and Guevara Noubir.
\newblock {Security of GPS/INS Based On-road Location Tracking Systems}.
\newblock In {\em {2019 IEEE Symposium on Security and Privacy (S\&P)}}, pages
  587--601, 2019.

\bibitem{Nishida2019-ue}
Erika Nishida.
\newblock {O brasileiro gosta mesmo {\'e} de monitorar o c{\^o}njuge via
  stalkerware}.
\newblock
  \url{https://gizmodo.uol.com.br/brasileiro-monitorar-conjuge-stalkerware/},
  October 2019.
\newblock Accessed: 2019-10-16.

\bibitem{Oas2019-eq}
{OAS}.
\newblock {IACHR} expresses deep concern over alarming prevalence of
  gender-based killings of women in brazil.
\newblock
  \url{http://www.oas.org/en/iachr/media_center/PReleases/2019/024.asp},
  February 2019.
\newblock Accessed: 2019-10-16.

\bibitem{Oliveira2017-uj}
Daniela Oliveira, Harold Rocha, Huizi Yang, Donovan Ellis, Sandeep Dommaraju,
  Melis Muradoglu, Devon Weir, Adam Soliman, Tian Lin, and Natalie Ebner.
\newblock {Dissecting Spear Phishing Emails for Older vs Young Adults: On the
  Interplay of Weapons of Influence and Life Domains in Predicting
  Susceptibility to Phishing}.
\newblock In {\em {Proceedings of the 2017 CHI Conference on Human Factors in
  Computing Systems}}, CHI '17, pages 6412--6424, New York, NY, USA, 2017. ACM.

\bibitem{PagBrasil2019-or}
{PagBrasil}.
\newblock {Mobile in Brazil: Usage Stats and User Profile}.
\newblock \url{https://www.pagbrasil.com/news/mobile-in-brazil/}, March 2019.
\newblock Accessed: 2019-12-7.

\bibitem{Peixoto2017-kv}
Danielle~Faria Peixoto.
\newblock A espacialidade da viol{\^e}ncia contra a mulher: Um estudo de caso
  na favela de rio das pedras ({RJ}).
\newblock {\em Women's Worlds Congress}, 2017.

\bibitem{Redmiles2019-bq}
E.~M. Redmiles.
\newblock {"Should I Worry?" A Cross-Cultural Examination of Account Security
  Incident Response}.
\newblock In {\em {2019 IEEE Symposium on Security and Privacy (S\&P)}}, pages
  920--934, May 2019.

\bibitem{Saboia2016-bc}
Fernanda Saboia.
\newblock {The Rise of WhatsApp in Brazil Is About More than Just Messaging}.
\newblock {\em Harvard Business Review}, April 2016.

\bibitem{Saboia2020-kw}
Gabriel Sab{\'o}ia.
\newblock {Sem internet, estudantes de favelas n{\~a}o conseguem se preparar
  para o Enem}.
\newblock
  \url{https://educacao.uol.com.br/noticias/2020/04/28/sem-internet-estudantes-de-favelas-sofrem-com-preparacao-online-para-enem.htm},
  April 2020.
\newblock Accessed: 2020-5-26.

\bibitem{Santiago2019-in}
Marisa~Antunes Santiago, Hebe~Signorini Gon{\c c}alves, and
  Cristiane~Brand{\~a}o Augusto.
\newblock {Mar{\'e} de Mulheres: Reflex{\~o}es sobre a Justi{\c c}a para
  Mulheres em Situa{\c c}{\~a}o de Violencia Numa Favela Carioca}.
\newblock {\em ex aequo - Revista da Associa{\c c}{\~a}o Portuguesa de Estudos
  sobre as Mulheres}, (40), December 2019.

\bibitem{Santiago_Navarro_F2016-dr}
Renata~Bessi Santiago Navarro~F.
\newblock {Terrorized by Police Raids and Mass Displacement, Rio Prepares for
  Olympics}.
\newblock
  \url{https://truthout.org/articles/terrorized-by-police-raids-and-mass-displacement-rio-prepares-for-olympics/},
  July 2016.
\newblock Accessed: 2020-5-16.

\bibitem{Waiselfisz2015-zs}
Julio~Jacobo Waiselfisz.
\newblock Mapa da viol{\^e}ncia 2015: Homic{\'i}dio de mulheres no brasil.
\newblock Technical report, ONU Mulher, OMS/OPAS, Secretaria de Políticas para
  as Mulheres, 2015.

\bibitem{Woodlock2017-hq}
Delanie Woodlock.
\newblock The abuse of technology in domestic violence and stalking.
\newblock {\em Violence Against Women}, 23, 2017.

\end{thebibliography}

\end{document}